\documentstyle[12pt]{article}
\def\NPB#1#2#3{Nucl. Phys.  {\bf B#1}  (19#2)  #3}
\def\PLB#1#2#3{Phys. Lett. {\bf B#1} (19#2) #3}

\def\PRD#1#2#3{Phys. Rev. {\bf C#1} (19#2) #3}
\def\PRL#1#2#3{Phys. Rev. Lett. {\bf#1} (19#2) #3}

\def\ARNP#1#2#3{Ann. Rev. Nucl. Part. Sci. {\bf#1} (19#2) #3}

\def\la{\lambda}
\def\f{\frac}
\def\ne{\nu_{e}}
\def\nm{\nu_{\mu}}

\def\sq{\sqrt{2}}

\def\beq{\begin{equation}}
\def\eeq{\end{equation}}
\newcommand{\newc}{\newcommand}
\newc{\sm}{Standard Model}
\newc{\smd}{Standard Model}
\newc{\barr}{\begin{eqnarray}}
 \newc{\earr}{\end{eqnarray}}
\newc{\dac}{discrete anomaly cancellation }
\newc{\mup}{``$\mu$'' problem }
\newc{\eps}{\epsilon}
\def\gappeq{\mathrel{\rlap {\raise.5ex\hbox{$>$}}
{\lower.5ex\hbox{$\sim$}}}}
\def\lappeq{\mathrel{\rlap{\raise.5ex\hbox{$<$}}
{\lower.5ex\hbox{$\sim$}}}}

\begin{document}

\begin{flushright}

hep-ph/9509351 \\
FTUAM-95-31 \\
HD-HEP-95-29 \\
UOI-95-321 \\
\end{flushright}

\begin{center}
{\bf TEXTURES FOR
NEUTRINO MASS MATRICES
} \\
\vspace*{1.0cm}
{\bf G.K. Leontaris$^{a}$},
{\bf S. Lola$^{b}$},
{\bf C. Scheich$^{c}$} and
{\bf J. D. Vergados}$^{a}$
\end{center}
\vspace*{0.1cm}
\begin{center}
\begin{tabular}{l}
$^{a}$
{\small Theoretical Physics Division, Ioannina University, Ioannina,
Greece}\\
$^{b}${\small Institut f\"{u}r
Theoretische Physik, Univerisit\"at Heidelberg,}\\
{\small Philosophenweg 16, 69120 Heidelberg, Germany }\\
$^{c}${\small Departamento de F\'isica Te\'orica, Universidad Aut\'onoma
de Madrid,}\\ {\small 28049, Madrid, Spain}\\
\end{tabular}
\end{center}

\begin{center}
{\bf ABSTRACT}
\end{center}

\small{
We give a classification of heavy Majorana neutrino mass
matrices with up to three texture zeroes, assuming
the Dirac masses of the neutrinos to be of the same form
as the ones of the up-quarks in the five texture zero solutions for the
quark matrices. This is the case for many unified
and partially unified models.
We find that it is possible to have solutions which account for
the solar and atmospheric neutrino problems as well as the COBE
observations simultaneously, and we motivate the existence of such solutions
from symmetries.
}

\noindent
\rule[.1in]{15cm}{.002in}

\noindent



\thispagestyle{empty}
\setcounter{page}{0}
\vfill\eject

\section{Introduction}

Recently, there has been a lot of work on the origin of fermion masses
and mixing angles\cite{5e,7,IR,DLLRS,LLR,oth}.
The observed pattern  of masses and mixings can be explained,
if some structure in addition
to that of the Standard Model exists at higher scales.
Within the context of supersymmetry,  unification has had considerable
success in determining  the parameters of the  Standard Model \cite{1a}.
Besides the successful predictions of the  gauge couplings, the pattern
 and magnitude of spontaneous symmetry breaking at the electroweak
scale \cite{radi} and the $b$ -- $\tau$ unification, it was
also found that the fermion mixing angles and masses have values
consistent with the appearance of ``texture'' zeros in the mass
matrices \cite{4,5e}. Such textures indicate the existence
of additional symmetries beyond the Standard Model and together with the
hierarchical structure observed in the quark and lepton mass
matrices imply that an underlying family
symmetry (e.g a $U(1)$ family symmetry)
with breaking characterized by a small parameter,
$\lambda$, might exist \cite{IR}.
For an exact symmetry, only the third generation would be massive and all
mixing angles zero. However,
symmetry breaking terms gradually fill the mass matrices in
powers of $\lambda$, generating a hierarchy of mass scales and
mixing angles. Thus, a broken symmetry can explain
the ``texture'' zeros as well as the relative magnitude of the
non-zero elements.

In a previous work \cite{DLLRS,LLR},
the implications of such a scheme for
neutrino masses and mixing angles in the case of the minimal
multiplet content of the MSSM (Minimal
Supersymmetric Standard Model), extended to include right-handed
neutrino components (plus the Standard Model singlet Higgses needed to
generate their masses and to break
the extended gauge family
symmetry) were considered.
Alternative
schemes have also been proposed
\cite{oth}).
In \cite{DLLRS,LLR} right-handed fields
got Majorana masses from a term of the form
$\nu_R\nu_R\Sigma$  where $\Sigma$ is a $SU(3)\otimes
SU(2)\otimes U(1)$ invariant Higgs scalar field with $I_W=0$ and
$\nu_R$ is a right-handed neutrino.  In many
models $\Sigma$ is not an elementary field, but
a combination of scalar fields $\Sigma = \tilde{\bar{\nu}}_R
\tilde{\bar{\nu}}_R$, where $\tilde{\bar{\nu}}_R$ is the scalar
component of a right-handed antineutrino supermultiplet,
${\bar\nu}_R$ \cite{su3}.
It was
found that, up to a discrete ambiguity,
the Majorana mass matrix is determined
and that,
unlike previous assumptions
for this matrix  \cite{7b}
a large splitting between the
entries exists.

The main conclusions of this work
were:

(i) The heaviest neutrino has a mass
$(0.4-4)$ eV for a top quark of $O(200)$ GeV,
thus being of the right size for structure formation \cite{hotdark}.

(ii) The light neutrinos have masses and mixing angles of the
magnitude needed to explain solar neutrino oscillations
\cite{solar}, \cite{rev},
in the small mixing angle region of the MSW effect.

(iii) In the simplest scheme, described in \cite{DLLRS},
it was not possible to
obtain large mixing angles,
without fine
tuning of the Yukawa couplings.
Such a large mixing is required for
a vacuum solution of the solar
neutrino problem as well as the
neutrino oscillation solution to the atmospheric
neutrino problem \cite{atmo}.
In \cite{LLR} it was found however that
in order to obtain  neutrino masses in a phenomenologically
interesting region while retaining bottom-tau unification,
in the small tan$\beta$ regime, large mixing in the
$\mu-\tau$ charged leptonic sector has to occur
\footnote{
The distortions to the bottom-tau unification
would appear as an implication of the structure emerging
from the $U(1)$ symmetry, if the right handed neutrinos have Yukawa
couplings of the same order as the up quarks, thus affecting the
radiative corrections in the model in the small tan$\beta$ regime
\cite{VB}.
An alternative solution arises in
a sub-class of
models where the Dirac-type Yukawa coupling of
the neutrino  is very suppressed \cite{dimp}.
In the large tan$\beta$ regime due to the infrared fixed
point for the bottom coupling, which is described by analytical
expressions in \cite{FLL},
the effect of the neutrino
coupling to bottom-tau unification is negligible.}.
This mixing can in principle appear also in the large tan$\beta$
regime and in particular in a subclass of the textures of
\cite{DLLRS}, when dropping a residual $Z_2$ symmetry.
Still, for a single $\Sigma$ field, the eigenvalues of
the light Majorana mass exhibit large splittings. Therefore,
although
we had been able to obtain two classes of solutions where
we address either the solar and the atmospheric neutrino problem,
or the solar neutrino problem and the COBE data,
it was not possible to solve
all three problems simultaneously.

However, in principle
there is no reason why this particular conclusion
in the simplest extension of the Standard Model,
should apply in the case of a more complicated symmetry
or with more than one pair of singlet fields,
$\Sigma ,{\bar\Sigma}$  present in the theory.
Since in such a case there are many possible patterns,
instead of making a complete search based on symmetries,
we follow the opposite
strategy, that is: (i) we assume the very large class of models from
underlying unified models (such as strings and grand unified
theories (GUTs)) or partially unified models that fix the
neutrino Dirac mass matrix to be proportional to the $u$-quark
mass matrix. (ii) we then study all possible
Majorana neutrino mass matrices with three exact and
an arbitrary number of phenomenological texture zeros.
It is clear that we are
looking for solutions with at least one large mixing angle
(to explain the atmospheric neutrino deficit)
and nearly degenerate
neutrino mass eigenvalues. (iii) We then motivate
these phenomenological solutions from symmetries.
As a result we find only a small number of possible Majorana
neutrino mass matrices. This gives a constraint on the underlying
theory in terms of necessary couplings.

We will start by briefly reviewing the experimental
limits on neutrino masses and mixing angles, in section 2.
In section 3 we will
discuss the whole framework of unification and mass matrices. In
section 4 we then give a first example of how we
find solutions with exact texture zeroes that allow large
mixing angles and nearly degenerate mass eigenvalues. The complete
results of this analysis are tabulated in the appendix.
These findings at high scales are then confronted with the low
energy requirements for such textures in section 5. Here a
classification of phenomenological texture zeroes is
given. Section 6 addresses the derivation of such textures from
underlying $U(1)$ symmetries.
The connection
of the textures at high and low scales
 via renormalization is discussed in section 7.
The conclusions are given in section 8.
Finally, the complete
approach of finding textures is summarized in
the appendix.

\section{Neutrino Phenomenology}

Various recent data, confirmed by many experimental groups, may be
explained if the neutrinos have non-vanishing masses. In such a case,
the neutral leptons produced in weak interactions are
in general not stationary. The weak eigenstates are linear
combinations of the neutrino mass eigenstates, implying
neutrino mixing. Although the Standard Model theory does not
give masses to the neutrinos, most extensions of the Standard Model
predict small masses and mixing. Before discussing such extensions,
we first review the experimental situation and give some
indication of possible explanations.

If the neutrinos are light, neutrino oscillation experiments are
the best candidates to measure the small mass differences
$\delta {m^2}$ (from $1eV^2$  down to $10^{-10} eV^2)$.
Furthermore, neutrino oscillations may explain the solar neutrino
problem, i.e.  the apparent reduction of the $\nu_e$ flux at earth
compared to that predicted by the standard solar model \cite{bah} (SSM).
If the neutrino mixing is small,
the mechanism of neutrino oscillations is not effective.
Nevertheless, under the conditions of high density
encountered in the sun's interior, the oscillation can be enhanced from
the MSW effect \cite{solar}, since small mixing angles can be
converted into large effective mixing angles, due to resonant scattering
of $\nu_e$ neutrinos by electrons.
The data from the solar neutrino experiments can thus be described either
by assuming resonant transitions (MSW-effect) or vacuum oscillations.
These two possibilities require the following ranges for masses and
mixing angles:

a) The small mixing angle solution for the MSW effect requires
\begin{eqnarray}
\delta m^2_{\nu_e\nu_{\alpha}}&\approx &(0.6-1.2)\times 10^{-5}
\; {\rm eV^2} \label{eq:m1}\\
sin^22\theta_{\alpha e}&\approx &(0.6-1.4)\times 10^{-2}\ .
\label{msw}
\end{eqnarray}

b) Vacuum oscillations can solve the solar neutrino puzzle if
\begin{eqnarray}
\delta m^2_{\nu_e\nu_{\alpha}}&\approx &(0.5-1.1)\times 10^{-10}
\; {\rm eV^2}\\
sin^22\theta_{\alpha e}&\ge  &0.75\ ,
\end{eqnarray}
where $\alpha$ is $\mu$ or $\tau$.
The most natural solution  in unified models is obtained through
the MSW-- mass and mixing angle ranges. This solution in particular
requires a light neutrino Majorana mass of the order
\beq
m_{\odot}\approx \sqrt{\delta m^2}\approx 3.0\times 10^{-3}eV\ ,
\eeq
as already given in (\ref{eq:m1}).
Such ultra light masses can be generated effectively in GUT's \cite{LAN}
and SUSY -- GUT's \cite{ROS} by the well known `see--saw' mechanism
\cite{HRR}.

The atmospheric neutrino problem may be explained in
the case that a large mixing and small mass splitting involving
the muon neutrino exists
\cite{atmo,fuk}.
Taking into account the
bounds from accelerator
and reactor disappearance
experiments one finds that
for $\nu_{e}-\nu_{\mu}$ or
$\nu_{\tau}-\nu_{\mu}$ oscillations
\beq
\delta m^2_{\nu_{\alpha}\nu_{\mu}} \leq
10^{-2} \; {\rm eV}^{2}
\label{at1}
\eeq
\beq
sin^22\theta_{\mu \alpha} \geq  0.51-0.6
\label{at2}
\eeq
where $\alpha$ stands for $e,\tau$ and in (\ref{at2}) the larger
lower limit for $sin^22\theta_{\mu \alpha}$ refers to $\nu_{\mu}-
\nu_{\tau}$ oscillations.
Finally we have already mentioned that neutrinos are possible candidates
for structure formation provided they have a mass of order $O({\rm eV})$.
This value is consistent (with a small margin according to some measurements)
with the bounds from neutrinoless double beta ($\beta\beta_{0\nu}$) decay.
In terms of the neutrino masses and mixing angles, the relevant
 $\beta\beta_{0\nu}$ measurable quantity
can be written as
\beq
|<{ m}_{\nu_e}>|=|\sum_i^3 (U_{ei}) ^2m_{\nu_i}e^{i\lambda_i}|
 \leq 1 eV\ , \label{eq:beta}
\eeq
where $e^{i \lambda_i}$ is the CP-parity of the $i^{th}$ neutrino,
while $U_{ei}$ are the elements of the unitary transformation
 relating the weak and mass neutrino eigenstates.

What are the possible theoretical solutions that are consistent
with this data?
Only a partial solution of all these problems
may  be obtained easily in simple models with a general
hierarchical pattern of neutrino masses. In most cases
it is possible to obtain a solution to the solar neutrino problem
with $m_{\nu_1}\ll m_{\nu_2}\approx m_{\odot}$, and
$m_{\nu_3}\sim O(1)eV$ to interpret the COBE data.
Then, the $\beta\beta_{0\nu}$ bound is satisfied due
to the smallness of the $U_{e3}$ mixing element
predicted by the theory. Indeed, assuming the above
hierarchy, the quantity $<{m}_{\nu_e}>$ may be approximated
by $\mid U_{e3}\mid ^2m_{\nu_3}$.
Due to the assumed hierarchy, it follows easily that
oscillation experiments are sensitive only to $\delta m_{13}^2$,
$\delta m_{23}^2$, since oscillations related
to $\delta m_{12}^2$ are too rapid. Thus the formula for the
oscillations $P(\nu_e\rightarrow \nu_e)$ \cite{cheng} is simplified
to
\begin{equation}
P(\nu_e\rightarrow \nu_e) = 1 - 4 {\mid U_{e3} \mid}^2
  (1 - {\mid U_{e3} \mid}^2) sin^2 \left( \frac{\pi x}{\ell} \right)\ ,
\end{equation}
where several trigonometric identities and $m_1, m_2 \ll m_3$ are used.
Setting $\mid U_{e3} \mid = sin \Theta_{ee}$ this may be rewritten as
\begin{equation}
P(\nu_e\rightarrow \nu_e) = 1 - sin^2 \Theta_{ee} sin^2
 \left( \frac{\pi x}{\ell} \right) \ .
\end{equation}
Taking  $sin^2\Theta_{ee} = 0.2$,  we find
$\mid U_{e3}\mid \approx  0.23$. This in turn would imply i.e $m_{\nu_3}
\approx 18.9$ eV for $<{ m}_{\nu_e}>\approx 1$ eV from above. Obviously,
the
atmospheric neutrino data does not fit in the above scenario.

It thus appears that the experimental data requires to have nearly
degenerate mass eigenstates $m_{\nu_i}\approx m_0$, $i=1,2,3$
\cite{psm}, since: First of all structure
formation in the Universe and the COBE data
requires $\sum_{i}m_i \approx 3$ eV.
 This sets the scale of the
masses. The data from the atmospheric and solar neutrino experiments
forces the involved masses to be very similar. In this case the
$\beta\beta_{0\nu}$ bound may be
respected due to mutual cancellations in (\ref{eq:beta})
by opposite CP--phases $e^{i\lambda_i}$.
Introducing an average mass $m_0$ one finds that
\begin{eqnarray} & &
\delta m_{12}^2\approx 2m_0 \mid m_2-m_1\mid \approx 10^{-5} eV^2 \\
& &
\delta m_{23}^2\approx 2m_0 \mid m_2-m_3\mid \leq 10^{-2} eV^2 \\
& &
<{ m}_{\nu_e}>\approx m_0 \sum_{i=1}^3 (U_{ei}) ^2e^{\imath\lambda_i}
 \leq 1 eV\ .
\end{eqnarray}
With the mentioned mutual cancellations,  $m_0\approx (1-2) eV$ can be
consistent with all data.
Our aim in the present paper is to use this observation and
constraints from the low energy theory, in order to determine
the optimal Majorana mass matrices with zero textures,
for a wide class of theories. We consider the
cases with {\it a) }hierarchical light-neutrino masses
(partial solutions) and with
{\it b)}
nearly degenerate neutrino mass eigenstates
of order $O({\rm eV})$ (complete solutions).

The necessary mixing may occur in two ways:
Either purely from the neutrino
sector of the theory, or
by the charged lepton mixing.
In the second case the mixing is typically too small
to have any impact on the atmospheric
neutrino problem, but may still account
for the solar neutrino problem.
In the former case, the mixing may be such as to
account for both deficits. One, of course, can consider
mixing in both sectors.
In this paper, we will search systematically for solutions with one large
mixing
angle, stemming from the need
to accommodate the atmospheric neutrino data from the beginning.
Such an origin of a large mixing actually seems to be the most
reasonable case. We also chose
the small mixing angle solution in order to address the solar
neutrino puzzle. We are not discussing the possibility of a second large
angle (as required for the vacuum oscillation solution for the
solar neutrinos), since the analysis will be more complicated.
Therefore we have two possibilities:

1) The solar neutrino problem is resolved by
$\nu_e-\nu_{\mu}$ oscillations and the atmospheric
neutrino problem by
$\nu_{\mu}-\nu_{\tau}$ oscillations.
For this possibility to be viable,
we need a large mixing
angle, in the 2-3 entries.

2)The solar neutrino problem is resolved by
$\nu_{e}-\nu_{\tau}$
oscillations and the atmospheric
neutrino problem by
$\nu_e-\nu_{\mu}$
oscillations.
In this case the large angle should be in the
1-2 entries.

\section{Unification and mass matrices}

In this section we discuss how predictions for mass matrices
arise in unified theories. We will start by resuming the discussion of
\cite{7,IR} for quark and lepton mass matrices
and show how this extends to
include neutrino mass matrices, if
we assume certain unifications.
So far, there has been a lot of progress in the construction of
viable string theories. Although many models seem to have fundamental
problems in resembling the standard model at low energies, most of them
can not be totally excluded,
because of their very complex structure \cite{diss}.
On the other hand there exist many more models, which have not been
studied at all. Therefore one is very interested in having an
additional criterion to distinguish between all these possibilities and
to single out those that may lead to the Standard Model at low energies.

The idea in \cite{IR} was
(instead of
taking specific models and studying their parameters),
to use
additional discrete symmetries, which
appear vastly in many string
models, and study their implications for fermion mass
patterns.
Discrete symmetries lead
quite naturally to hierarchies of parameters (Yukawa couplings etc.)
and therefore to predictions that are largely independent of specific
numerical values.
A certain model is realistic, only if
it possesses such a hierarchical structure,
without any fine-tuning.
Indeed, it is a general
observation that a string
model exhibits all possible couplings allowed by the
discrete symmetries, which are typically of $O(1)$\footnote{Here
we are referring to models close to e.g. the conformal point and
not to large moduli VEV (vacuum expectation value)
cases.} and
there are hardly any {\it accidental} zeroes
for the values of the Yukawa couplings.

To see how symmetries may imply a hierarchical
pattern of Yukawa couplings (and therefore masses),
one can look at the mass matrices for a two quark doublet, assuming an
additional $U(1)$ symmetry to the standard model
gauge group, under which
$t_{L,R}, b_{L,R}$ have charge $\alpha_{1}$ and
$c_{L,R}, s_{L,R}$ have charge $\alpha_{2}$.
The form of the actual mass matrix depends also on the
transformation of the Higgs fields $H_1$, $H_2$
which give masses to the fermions. Taking these
charges to be $-2 \alpha_{1}$, only the $t$- and $b$-quarks
acquire a mass in the electroweak
breaking and the up or down
quark mass matrix has only one non-zero element.
However, if the $U(1)$ symmetry is broken
to a discrete symmetry
$Z_{N}$ by the VEV of a
field $\theta$ with charge $-1$,
there are several mechanisms that give structure to the mass
matrices:

1) {\it Higher dimensional operators}
$q_Lq_RH_i\frac{\theta^n}{M^n}$, where $M$ is the scale where these
terms are created. Then in general the mass matrices are of the form
\begin{equation}
\left( \begin{array}{cc} \lambda^2 & \lambda \\
                         \lambda   & 1       \end{array} \right)\ ,
\end{equation}
where $\lambda=\frac{<\theta>}{M}$
and a hierarchy in terms of $\lambda$ arises.

2) {\it Mixing of the coupling Higgs field}

3) {\it Mixing of the coupling matter fields}

Thus additional symmetries, together with stages
of spontaneous symmetry breaking, allow for a natural explanation
of hierarchies of masses.
This actually
makes use of the huge discrete symmetry groups appearing in string
models, therefore one may hope that the patterns of mass matrices
may help to determine the underlying discrete symmetries and in turn to give
restrictions on a possible fundamental string theory.

Let us note here that there are {\it two types} of texture
zeroes: {\it exact} and {\it phenomenological}.
The first type is a zero implied by a symmetry.
The second type, relaxes the zeroes
in a way that does not change the hierarchical structure of a given
matrix.
If we are assuming a
fundamental theory that has only certain couplings at the tree level, like a
pure GUT theory,
it is clear from the start that we deal with
exact zeroes.
On the other hand,
zeroes incorporating mechanisms 1 to 3 to create
entries in the mass matrices seem bound to be phenomenological.

After giving the underlying idea of creating hierarchies of
couplings  and therefore masses, let us briefly quote the
``phenomenological'' results of Ramond,
Roberts and Ross \cite{7}, which we will need
for the discussion that follows.
Here, the authors looked for
parameterizations of symmetric quark mass matrices
in terms of
possible texture zeroes and a
hierarchical parameter $\lambda$ that is in accordance with
experiment. Those structures would have then to be explained
by symmetries of the underlying theory.
Although a pattern of zeroes in one single
mass matrix does not have a meaning on its own (because of the
possible redefinitions of the quark fields), it has a meaning
for the up and
down quark mass matrices together.
Therefore, one encounters a {\it relative
structure} (e.g. one matrix may be made diagonal by redefinitions,
but the texture zero structure determines now the other matrix.)
A complete study of 5 and 6 texture zeroes
in the two mass matrices has been carried out along these lines.
Studying systematically all the possible cases and taking
into account the running of the renormalization group equations
between the unification scale $M_X$, where the texture zeroes
are assumed, and $M_W$,  five realistic pairs of texture
zero patterns for the quark mass matrices were found.
These appear in table
\ref{table:maj}.

\begin{table}
\centering
\begin{tabular}{|c|c|c|} \hline
Solution & $Y_{u}, m^D_{\nu}$ & $Y_{d}$
\\ \hline
1 & $\left(
\begin{array}{ccc}
0 & \sq \la^{6} & 0 \\
\sq \la^{6} & \la^{4} & 0 \\
0 & 0 & 1
\end{array}
\right)$ &
$\left(
\begin{array}{ccc}
0  & 2 \la^{4}  & 0 \\
2 \la^{4}  & 2 \la^{3} & 4 \la^3 \\
0  &  4 \la^3 & 1
\end{array}
\right)$
\\ \hline

2 & $\left(
\begin{array}{ccc}
0 &  \la^{6} & 0 \\
 \la^{6} & 0 & \la^2 \\
0 & \la^2 & 1
\end{array}
\right)$ &
$\left(
\begin{array}{ccc}
0  & 2 \la^{4}  & 0 \\
2 \la^{4}  & 2 \la^{3} & 2\la^3 \\
0  &  2\la^3 & 1
\end{array}
\right)$
\\ \hline

3 & $\left(
\begin{array}{ccc}
0 & 0 & \sq \la^4 \\
0 & \la^{4} & 0 \\
\sq \la^4 & 0 & 1
\end{array}
\right)$ &
$\left(
\begin{array}{ccc}
0  & 2 \la^{4}  & 0 \\
2 \la^{4}  & 2 \la^{3} & 4 \la^3 \\
0  &  4 \la^3 & 1
\end{array}
\right)$
\\ \hline

4 & $\left(
\begin{array}{ccc}
0 & \sq \la^{6} & 0 \\
\sq \la^{6} & \sqrt{3}\la^{4} & \la^2 \\
0 & \la^2 & 1
\end{array}
\right)$ &
$\left(
\begin{array}{ccc}
0  & 2 \la^{4}  & 0 \\
2 \la^{4}  & 2 \la^{3} & 0 \\
0  &  0 & 1
\end{array}
\right)$
\\ \hline

5 & $\left(
\begin{array}{ccc}
0 & 0 & \la^4 \\
0 & \sq \la^{4} & \la^2/\sq \\
\la^4 & \la^2/\sq & 1
\end{array}
\right)$ &

$\left(
\begin{array}{ccc}

0  & 2 \la^{4}  & 0 \\
2 \la^{4}  & 2 \la^{3} & 0 \\
0  &  0 & 1
\end{array}
\right)$  \\
\hline
\end{tabular}
\caption{
Approximate forms for the symmetric
textures.}
\label{table:maj}
\end{table}

In general the texture zero structure is not
unaffected by the running of the renormalization group.
Nevertheless the hierarchy structure is
preserved, indicating that the texture zeroes are kept at least
as {\it phenomenological} ones. We refer the discussion
to the following sections.

Let us now turn from the quark masses to
lepton and neutrino masses.
In the case of the neutrino masses an
additional complication, through Majorana masses, arises.
The experimentally relevant light neutrino mass matrix is
 given by
\begin{equation}
m^{eff}_{\nu}=m^D_{\nu}\cdot (M_{\nu_R})^{-1}\cdot m^{D\dagger}_{\nu}\ ,
\label{eq:meff}
\end{equation}
where $m_{\nu}^D$ is the Dirac neutrino mass matrix and $M_{\nu_R}$
the heavy Majorana neutrino mass matrix. $m_{\nu}^D$
and $M_{\nu_R}$ are completely generic, unless
we assume a unification
that makes the prediction
$m_{\nu}^D \sim m_u$.
There are two ways of reasoning such an assumption. Either by a
unified or by a partially unified theory. Such relations are then based on
a GUT or a string theory
\footnote{One should think that the (heterotic) string
theory is the prefered scenario, since it allows the solution of more
fundamental problems and also delivers a rich structure of discrete
symmetries which may serve to introduce zeroes and
hierarchical patterns in the
mass matrices.}.
For the GUT
theories, the gauge groups $E_6$, $SU(5)$ and $SO(10)$ allow
for an interesting phenomenology.
In the case of string gauge groups such as $E_6$ or
subgroups of the same rank (after Wilson line breaking) as well as
in the flipped version of $SU(5)$, interesting relations between
fermion masses also appear quite naturally.
Many of these models
contain multiplets that allow for the same structure of the
$u$-quark mass matrix $m_u$ and the Dirac neutrino mass matrix
$m_{\nu}^D$ \footnote{This however has to be taken with a
grain of salt, since
in the (very interesting) case of Wilson line breaking of $E_6$ in
a heterotic string theory this structure is not perpetuated to the
broken theory \cite{su3,diss}.}.

Concering the heavy Majorana neutrino mass matrices,
we will demand hermitian matrices
$M_{\nu_R}$. This
will enable us to classify the general three exact and any number of
phenomenological texture zero solutions
and to give some insight into solutions with less
exact texture zeroes.

Before passing to specific examples we
have to discuss the lepton mass matrices,
since their diagonalizing matrix enters in the mixing matrix of
the charged leptonic currents. In complete analogy to the quark
currents the leptonic (KM) mixing matrix is
\beq
V_{tot} =
V_{\ell}V_{\nu}^{\dagger}\ ,
\label{eq:mix}
\eeq
where $V_{\ell}$ diagonalizes the charged lepton mass matrix, while
$V_{\nu}$ diagonalizes the light neutrino mass matrix.
Instead of making a specific assumption for the lepton mass
matrix (or equivalently) the associated Yukawa couplings,
we just treat them
as parameters. We will only apply the observation that the
charged lepton
hierarchical structure usually does not lead to large
mixings.
E.g. the ansatz in \cite{giudice}, which is case 3 in table \ref{table:maj}
for
the quark masses and $m_e \approx m_d$ up to a numerical factor,
has been studied
extensively
and it was found \cite{george}
that
the mixing matrix due to
charged current interactions,
$V_{\ell}$ is given by
\beq
V_{\ell} =
\left (
\begin{array}{ccc}
1 & s_{3} & -s_{2} \\
-s_{3} & 1 & s_{1} \\
s_{2} & -s_{1} & 1
\end{array}
\right)
\eeq
where
the parameters
$s_{1,2,3}$ are
determined in terms of
fermion mass ratios.
For this ansatz
\begin{eqnarray}
s_{3} & = & 6.9 \cdot 10^{-2}
\nonumber \\
s_{1} & = & 3.95 \cdot 10^{-2}
\nonumber \\
s_{2} & \sim &  \frac{m_{c}}{m_{t}}\sim 10^{-2}\ \ .
\end{eqnarray}
(Here we omitted possible complex phases, since they should be irrelevant
when just discussing the mixing alone.) This indicates that while the
$e-\tau$ mixing is too small to have any importance
for  the MSW effect, the $e-\mu$ mixing is sufficiently large.
The total mixing matrix for the neutrinos is given by (\ref{eq:mix}).
This indicates that in this ansatz, even if MSW oscillations can not be
generated only via $V_{\nu}$,
including the mixing coming from
the charged current interactions
may lead to a solution.

\section{The form of the Majorana mass matrix - a first example}

In this section we will consider a first example of a model
with exact texture zeroes, which
potentially allows the consistent incorporation of all experimental data.
Here we will study the case with a strong mixing in the
2-3 entries of the effective neutrino
mass matrix ($\nu_{\mu}-\nu_{\tau}$ mixing).
This will then enable a solution of the atmospheric neutrino problem.
For simplicity,
we assume that MSW oscillations are generated due to the mixing
that arises from the charged current interactions.
Furthermore we want to have nearly degenerate masses. To
simplify the analysis, we take the 1-2 and 1-3 mixing angles to be zero
in this simple example. The Dirac mass matrix is taken to be given by
the Giudice ansatz.

In order to identify the possible forms of the heavy Majorana mass
matrix, we start from an effective light mass matrix with a strong
 mixing. We then investigate which form of the heavy
Majorana mass matrix is compatible with the specific form of the neutrino
Dirac mass matrix. This is the procedure we will follow in
the appendix, in order to obtain a full range of viable patterns for the
heavy Majorana mass matrix. There, we also discuss the issue of the
complex phases involved in all the mixings.

We invert (\ref{eq:meff})
\begin{equation}
m_{eff}^{-1} = (m_{\nu}^{D\dagger})^{ -1}
\cdot (M_{\nu_R})\cdot (m_{\nu}^{D})^{-1}\ ,
\label{eq:inv}
\end{equation}
to get
\begin{equation}
M_{\nu_R} = m_{\nu}^{D\dagger}\cdot m_{eff}^{-1}\cdot m_{\nu}^D\ \ .
\label{eq:done}
\end{equation}
where $m_{eff}^{-1\ diag}$ is given by
\beq
m^{-1\ diag}_{eff} =
\left (
\begin{array}{ccc}
\frac{1}{m_{1}} & 0 & 0 \\
0 & \frac{1}{m_{2}} & 0 \\
0 & 0 & \frac{1}{m_{3}}
\end{array}
\right)
\eeq
With the mixing matrix
\beq
V_{\nu} = \left
(\begin{array}{ccc}
1 & 0 & 0 \\
0 & c_{1} & -s_{1} \\
0 & s_{1} & c_{1}
\end{array}
\right)\ .
\eeq
$m^{-1}_{eff} = V_{\nu}m_{eff}^{-1diag}V_{\nu}^T$ has the form
\beq
m_{eff}^{-1} = \left (
\begin{array}{ccc}
\frac{1}{m_{1}} & 0 & 0 \\
0 & \frac{c^2_1}{m_2}+\frac{s_1^2}{m_3} &
c_1s_1(\frac{1}{m_2}-\frac{1}{m_3})
\\
0 &
c_1s_1(\frac{1}{m_2}-\frac{1}{m_3})
& \frac{c_1^2}{m_3}+\frac{s^2_1}{m_2}
\end{array}
\right)
\;
\equiv
\;
\left (
\begin{array}{ccc}
a & 0 & 0 \\
0 & b & d \\
0 & d & c
\end{array}
\right)\ .
\label{eq:form}
\eeq
Identifying the entries
gives
\begin{eqnarray}
\sin^{2}2\theta_{1} & = &
\frac{4 d^{2}}{(m_2^{-1}-m_3^{-1})^2}
\nonumber \\
m_1^{-1} & = & a
\nonumber \\
m_2^{-1} & = &
\frac{b}{2}
+ \frac{c}{2} + \frac{1}{2}
\sqrt{
b^{2}-2bc+c^2+4d^2}
\nonumber \\
m_3^{-1} & = &
\frac{b}{2}
+ \frac{c}{2} - \frac{1}{2}
\sqrt{
b^{2}-2bc+c^2+4d^2}\ ,
\end{eqnarray}
where $\theta_{1}$ is the
$\nu_{\mu}-\nu_{\tau}$ mixing angle.

The case of the absolute value of the three masses equal (i.e.
$m_1 = m_2$, $m_3 = -m_2$\footnote{The fact that these relative signs
between the masses are of fundamental importance will be discussed in
the appendix.}) is equivalent to
\beq
b = c = 0, \; \; \;
a = d\ .
\eeq
Therefore
\beq
\sin^{2}2\theta_{1} = 1, \; \; \;
\theta_1 = 45^{0}\ .
\eeq
The form of the heavy Majorana mass
matrix may then easily be found from (\ref{eq:done}).
For the Giudice ansatz\footnote{The reader should keep in mind that
this ansatz differs from the case 3 in table \ref{table:maj} by two factors of
$\sqrt{2}$. Therefore we take here $x$ instead of $\lambda$ to denote
the difference.},
where (after rescaling)
\beq
m_{\nu}^{D} =
\left (
\begin{array}{ccc}
0 & 0 & x \\
0 & x & 0 \\
x & 0 & 1
\end{array}
\right)\ ,
\eeq
we find that
\beq
\left (
\begin{array}{ccc}
x^{2}
(\frac{c_{1}^{2}}{m_{3}}
+ \frac{s_{1}^{2}}{m_{2}})
& x^{2}\frac{sin2\theta_{1}}{2}\left
(\frac{1}{m_2}-\frac{1}{m_3}\right)
& x
(\frac{c_{1}^{2}}{m_{3}}
+ \frac{s_{1}^{2}}{m_{2}}) \\
x^{2}\frac{sin2\theta_{1}}{2}\left
(\frac{1}{m_2}-\frac{1}{m_3}\right)
& x^{2}\left
(\frac{c_{1}^{2}}{m_{2}}
+ \frac{s_{1}^{2}}{m_{3}}
\right )
& x\frac{sin2\theta_{1}}{2}\left
(\frac{1}{m_2}-\frac{1}{m_3}\right)
\\
x^{2}\left
(\frac{c_{1}^{2}}{m_{3}}
+ \frac{s_{1}^{2}}{m_{2}}
\right )
& x\frac{sin2\theta_{1}}{2}\left
(\frac{1}{m_2}-\frac{1}{m_3}\right)
& \frac{x^2}{m_1} +
\frac{c_1^2}{m_3}+
\frac{s_1^{2}}{m_{2}}
\end{array}
\right)\ .
\eeq
For the above values of the three light masses this becomes
\beq
M_{\nu_R}\ = \
\left (
\begin{array}{ccc}
0 & M_{N} x & 0 \\
M_{N} x & 0 & M_{N}  \\
0 & M_{N}   &  M_{N} x
\end{array}
\right)\ ,
\label{eq:pattern}
\eeq
where $M_{N} = xd \approx 10^{11}-10^{13}$  GeV.
Thus, we see that in this example the degeneracy of all three masses
and one large  mixing angle is consistent and may be understood in terms
of texture zeroes  of the heavy Majorana neutrino mass matrix $M_{\nu_R}$  at
the scale $M_X$. We stress that such three texture zero solutions are maximal.
More zeroes normally\footnote{See also the discussion about this point
in the appendix.} imply vanishing determinant and less texture zeroes
are less predictive.

If we only have $c = 0$, then the heavy Majorana mass matrix becomes
\beq
M_{\nu_R}\ = \
\left (
\begin{array}{ccc}
0 &M_{N} x & 0 \\
M_{N} x & \frac bd M_N x & M_{N}  \\
0 & M_{N}   &\frac ad M_{N} x
\end{array}
\right)\ , \label{eq:heavy}
\eeq
where $M_N \approx xd$ and it possesses less texture zeroes than before.

For the systematic study of the three texture zero solutions, we
refer the reader to the the appendix. Here all possible cases of solutions
with at least one large mixing angle are given.

\section{Study of viable Majorana mass matrices}

In this section we will discuss mass matrices at the low energy
scale $M_W$. So far we studied exact texture zeroes of neutrino mass
matrices at the unification scale $M_X$. To investigate their
impact on mass matrices at $M_W$, one has to perform a renormalization
group analysis. As already mentioned, the exact texture zeroes
are in general not preserved. Nevertheless, the hierarchical
structure is kept. Or to say it in other words: {\it exact} texture
zeroes become {\it phenomenological} ones. We therefore want to study
such phenomenological zeroes at $M_W$ here and
 confront these solutions with
the preliminary solutions at $M_X$ in section 7.
At this stage we want to stress that
a discussion of phenomenological zeroes at $M_W$ immediately applies
to $M_X$ as well, the reason being the preservation of the hierarchical
structure by the renormalisation group running between the two scales.

We take the admissible Dirac mass matrices from table \ref{table:maj} and
study again solutions of (\ref{eq:inv}), assuming
one large angle to solve
the atmospheric neutrino and drop any further mixing in $m_{eff}^{-1}$.
We may then imagine the small mixing (needed for the solar neutrino
deficit)
to be due to the phenomenological character of the zeroes or to
reside in the charged lepton mixing. Therefore this approach is less
stringent than the one in section 4 and
the appendix, where the small mixing at $M_X$
was taken to be zero or physically trivial
for all three zero textures. Here we will
parameterize the small mixing in the appropriate way.

We start with an atmospheric neutrino mixing residing in the
2-3 submatrix. $m_{eff}^{-1}$ then takes the form (\ref{eq:form}), which
we use as a convenient parameterization.
\begin{table}
\centering
\begin{tabular}{|c|c|c|} \hline
Solution & $ M_{\nu_R}$ & Comments
\\ \hline
1a & $\left(
\begin{array}{ccc}
0 & 0 & \frac{d}{c}\sq \la^{6}  \\
0 & 2\frac{a}{c} \la^{12} & \f{d}{c}\la^{4}  \\
\f{d}{c}\sq\la^6 & \f{d}{c} \la^4 & 1
\end{array}
\right)$ &
for $b=0$
\\ \hline

1b & $\left(
\begin{array}{ccc}
2\frac{b}{d}\la^8 & \sq \frac{b}{d}\la^6 & \sq \la^{2}  \\
\sq \frac{b}{d}\la^6 & \frac{b}{d} \la^{4}+2\frac{a}{d}\la^8 & 1  \\
\sq\la^2 & 1 & 0
\end{array}
\right)$ &
for $c=0$
\\ \hline

2 & $\left(
\begin{array}{ccc}
0 & \f{d}{c}\la^8 & \frac{d}{c} \la^{6}  \\
\f{d}{c}\la^8 & \la^4 +\frac{a}{c}\la^{12}&
\la^2 [+\frac{d}{c}\la^4] \\
\f{d}{c}\la^6 & \la^2 [+\frac{d}{c}\la^4] & 1 [+2\frac{d}{c}\la^2]
\end{array}
\right)$ &
for $b=0$
\\ \hline

3a & $\left(
\begin{array}{ccc}
0 & \sq \la^4 & 0 \\
\sq \la^4 & \f{b}{d}\la^4 & 1 \\
0 & 1 &  2 \f{a}{d}\la^4
\end{array}
\right)$ & for $c=0$
\\ \hline

3b & $\left(
\begin{array}{ccc}
2\la^8 & \sq \frac{d}{c} \la^8 & \sq \la^4 \\
\sq \frac{d}{c} \la^8 & 0 & \frac{d}{c} \la^4 \\
\sq \la^4 & \frac{d}{c} \la^4 &  1+2 \f{a}{c}\la^8
\end{array}
\right)$ & for $b=0$
\\ \hline

4 & $\left(
\begin{array}{ccc}
0 & \frac{d}{c}\sq\la^8 & \f{d}{c}\sq\la^6 \\
\frac{d}{c}\sq\la^8 & \la^4 [+2\sqrt{3}\frac{d}{c}\la^6]
+2\frac{a}{c}\la^{12}
& \la^2 [+(1+\sqrt{3})\frac{d}{c}\la^4] \\
\f{d}{c}\sq\la^6 & \la^2 [+(1+\sqrt{3})\frac{d}{c}\la^4]
& 1 [+2\frac{d}{c}\la^2]
\end{array}
\right)$ &
for $b=0$
\\ \hline


5 & $\left(
\begin{array}{ccc}
0 & \la^6 & \f{1}{2}\la^4 \\
\la^6 & \sq \la^4 & \f{1+2\sq}{4}\la^2 [+\frac{b}{d}
\frac{1}{\sqrt{2}}\la^4] \\
\f{1}{2}\la^4 & \f{1+2\sq}{4}\la^2 [+\frac{b}{d}
\frac{1}{\sqrt{2}}\la^4] & 1 [+\frac{b}{d}\frac{\la^2}{2\sq}]+
\frac{a}{d}\frac{1}{\sq}\la^6
\end{array}
\right)$ &
for $c=0$ \\
\hline
\end{tabular}
\caption{The texture zero solutions of the Majorana mass
matrices associated with
each of the Dirac mass textures of table 1 with a large
mixing in the 2-3 submatrix. We present here cases where
either $b=0$ or $c=0$.
The non-leading powers are in brackets except for
the terms containing the parameter $a=\frac{1}{m_1}$.}
\label{table:maj2}
\end{table}
\begin{table}
\centering
\begin{tabular}{|c|c|c|} \hline
Solution & $ M_{\nu_R}$ & Comments
\\ \hline
1 & $\left(
\begin{array}{ccc}
0&0&{\sqrt{2}}\,d\,{\lambda^6} \\
0&2\,a\, {\lambda^{12}}&d\,{\lambda^4} \\
{\sqrt{2}}\,d\,{\lambda^6}&d\,{\lambda^4}&0
\end{array}
\right)$  &
for $b=c=0$
\\ \hline

2 & $\left(
\begin{array}{ccc}
0 & d\,{\lambda^8}&d\,{\lambda^6}  \\
d\,{\lambda^8}&a\,{\lambda^{12}}&d\,{\lambda^4} \\
d\,{\lambda^6}&d\,{\lambda^4}&2\,d\,{\lambda^2}
\end{array}
\right)$ &
for $b=c=0$
\\ \hline

3 & $\left(
\begin{array}{ccc}
0&{\sqrt{2}}\,d\,{\lambda^8}&0 \\
{\sqrt{2}}\,d\,{\lambda^8}&0&d\,{\lambda^4}\\
0&d\,{\lambda^4}&2\,a\,{\lambda^8}
\end{array}
\right)$ & for $b=c=0$
\\ \hline

4 & $\left(
\begin{array}{ccc}
0&{\sqrt{2}}\,d\,{\lambda^8}&{\sqrt{2}}\,d\, {\lambda^6} \\
{\sqrt{2}}\,d\,{\lambda^8}&2\,{\sqrt{3}}\,d\,
{\lambda^6} + 2\,a\,{\lambda^{12}}&
    d\,{\lambda^4} + {\sqrt{3}}\,d\,{\lambda^4}\\
{\sqrt{2}}\,d\,{\lambda^6}&d\,{\lambda^4} +
 {\sqrt{3}}\,d\,{\lambda^4}&2\,d\,{\lambda^2}
\end{array}
\right)$ &
for $b=c=0$
\\ \hline

5 & $\left(
\begin{array}{ccc}
0&{\sqrt{2}}\,d\,{\lambda^8}&
{{d\,{\lambda^6}}\over {{\sqrt{2}}}} \\
{\sqrt{2}}\,d\,{\lambda^8}&2\,d\,{\lambda^6}&
    {{d\,{\lambda^4}}\over 2} + {\sqrt{2}}\,d\,{\lambda^4} \\
{{d\,{\lambda^6}}\over {{\sqrt{2}}}}&
    {{d\,{\lambda^4}}\over 2} + {\sqrt{2}}\,d\,{\lambda^4}&
    {\sqrt{2}}\,d\,{\lambda^2} + a\,{\lambda^8}
\end{array}
\right)$ &
for $b=c=0$ \\
\hline
\end{tabular}
\caption{Cases as in table 2, but with $b=c=0$.}
\label{table:maj3}
\end{table}
\begin{table}
\centering
\begin{tabular}{|c|c|c|} \hline
Solution & $ M_{\nu_R}$ & Comments
\\ \hline
1 & $\left(
\begin{array}{ccc}
 0 & 2\frac{d}{c}\la^{12} & 0 \\
 2\frac{d}{c}\la^{12} & 2\sq\frac{d}{c}\la^{10} + 2\frac{a}{c}\la^{12} & 0 \\
 0 & 0 & 1
\end{array}
\right)$  &
for $b=0$
\\ \hline

2 & $\left(
\begin{array}{ccc}
 0 & \frac{d}{c}\la^{12} & 0  \\
 \frac{d}{c}\la^{12} & \la^4 + \frac{a}{c}\la^{12} &
 \la^2 + \frac{d}{c}\la^8 \\
 0 & \la^2 + \frac{d}{c}\la^8 & 1
\end{array}
\right)$ &
for $b=0$
\\ \hline

3 & $\left(
\begin{array}{ccc}
 2\la^8 & 0 & \sq\la^4 \\
 0 & 0 & \sq\frac{d}{c}\la^8\\
 \sq\la^4 & \sq\frac{d}{c}\la^8 & 1 + 2\frac{a}{c}\la^8
\end{array}
\right)$ & for $b=0$
\\ \hline

4 & $\left(
\begin{array}{ccc}
 0 & 2\frac{d}{c}\la^{12} & 0 \\
 2\frac{d}{c}\la^{12} & 2\sqrt{6}\frac{d}{c}\la^{10}+2\frac{a}{c}\la^{12} &
 \la^2 + \sq\frac{d}{c}\la^8 \\
 0 & \la^2 + \sq\frac{d}{c}\la^8 & 1
\end{array}
\right)$ &
for $b=0$
\\ \hline

5 & $\left(
\begin{array}{ccc}
 \la^8 & \frac{\la^6}{\sq} & \la^4 \\
 \frac{\la^6}{\sq} & \frac{\la^4}{2} & \frac{\la^2}{\sq} +
 \sq\frac{d}{c}\la^8 \\
 \la^4 & \frac{\la^2}{\sq} + \sq\frac{d}{c}\la^8 &
 1 + 2\frac{d}{c}\frac{\la^6}{\sq} + \frac{a}{c}\la^8
\end{array}
\right)$ &
for $b=0$ \\
\hline
\end{tabular}
\caption{The texture zero solutions of the Majorana mass matrices
associated with each of the Dirac mass textures of table 1 with a large
mixing in the 1-2 submatrix, for the examples with $b=0$. Only cases
for $b=0$ emerge and the solutions for $a=b=0$ follow immediately.}
\label{table:maj41}
\end{table}
The solutions (\ref{eq:done}) of
(\ref{eq:inv}) allowing for texture zeroes\footnote{These zeroes are
only of phenomenological type mainly due to the effect of the RG on
$m_{\nu}^D$.} in $M_{\nu_R}$ are given in tables \ref{table:maj2}
and \ref{table:maj3}.
Textures arising from a large mixing in the 1-2 submatrix
appear in table \ref{table:maj41}. Here $m_{eff}^{-1}$ takes a form
similar to (\ref{eq:form}), where the off diagonal elements, $d$, appear
in the 1-2 submatrix.

We now pass to a discussion of the phenomenology induced by the
forms of $m_{eff}^{-1}$ that have been quoted. We investigate the
case of a large mixing in the 2-3 submatrix. The case of
large mixing in the 1-2 submatrix is very much the same
and leads to analogous conclusions.

There are  two possibilities for texture zero solutions:
$b=0$ or $c=0$\footnote{The case $b=c=0$ e.g. has been already discussed
for the Dirac mass matrix pattern 3 of table 1
in section 4 and implies $\xi=1$,
is therefore in accordance with (\ref{eq:grenzen}).}
that follow.

\noindent (i) {\it c = 0}

Imposing this constraint onto (\ref{eq:form}) suggests a
 rewriting in terms of the parameter $\xi = -\frac{m_2}{m_3} > 0$.
Then
\beq
c_1 = \frac{1}{\sqrt{1+\xi}}, \; \; \; \; \; \;
s_1 = \frac{\sqrt{\xi}}{\sqrt{1+\xi}}\ ,
\label{eq:sin}
\eeq
\beq
m_{eff}^{-1} =
\left (
\begin{array}{ccc}
\frac{1}{m_1} & 0   &    0 \\
0 & \frac{1-\xi}{m_2} & \frac{\sqrt{\xi}}{m_2} \\
0 & \frac{\sqrt{\xi}}{m_2} &  0
\end{array} \right)
\label{eq:gros}
\eeq
and thus
\beq
sin^2 2\theta_1 = \frac{4\xi}{(1+\xi)^2}\ .
\label{eq:wink}
\eeq
The neutrino oscillation probabilities are given in terms
of the mixing matrix (where the origin of $\theta_e$ is undetermined,
as already said)
\beq
V_{tot} = V_{e}^{\dagger} V_{\nu} =
\left (
\begin{array}{ccc}
c_{e} & -s_{e} & 0 \\
s_{e} & c_{e} & 0 \\
0 & 0 & 1
\end{array} \right) \cdot
\left (
\begin{array}{ccc}
1 & 0 & 0 \\
0 & c_{1} & s_{1} \\
0 & -s_{1} & c_{1} \\
\end{array}
\right)
\label{eq:multi}
\eeq
\beq
V_{tot} =
\left (
\begin{array}{ccc}
c_{e} & -s_{e}c_1 & -s_{e}s_1 \\
s_{e} & c_{e}c_1 & c_{e}s_1 \\
0 & -s_1 & c_1
\end{array} \right) \ , \label{eq:mixer}
\eeq
where we take
\beq
s_{e} \approx
\sqrt{\frac{m_{e}}{m_{\mu}}} \approx 0.07, \; \; \; c_{e} \approx 1\ .
\label{eq:zahl}
\eeq
Such an ansatz for the charged leptons is most commonly used \cite{5e}.
Furthermore the {\it block} form (\ref{eq:multi}) seems appropriate to
accommodate the data, since (\ref{eq:m1}),(\ref{msw}) and (\ref{at1}),
(\ref{at2})
strongly suggest this. A more general ansatz is definitely more difficult
to handle.

It is now straightforward to calculate the oscillations
$P(\nu_{\alpha} \rightarrow \nu_{\beta})$ for (\ref{eq:mixer}), using
some identities and the general formula from \cite{cheng}.
We thus obtain
\beq
P(\nu_{\mu} \rightarrow \nu_{\tau}) =
c_e^2 \f{4\xi}{(1+\xi)^2} \sin^2
\f{m_2^2 (1/\xi^2-1)x}{4 E_{\nu}}
\eeq
\beq
P(\nu_{e} \rightarrow \nu_{\tau}) =
s_e^2 \f{4\xi}{(1+\xi)^2} \sin^2
\f{m_2^2 (1/\xi^2-1)x}{4 E_{\nu}}
\eeq
\begin{displaymath}
P(\nu_{e} \rightarrow \nu_{\mu}) =
\sin^2 2\theta_{e} \left[
\f{1}{(1+\xi)} \sin^2
\f{(m_2^2 -m_1^2) x} {4 E_{\nu}}
\right.
\end{displaymath}
\beq
+ \f{\xi}{(1+\xi)} \sin^2
\f{(m_3^2 -m_1^2) x}{4 E_{\nu}}
- \f{\xi}{(1+\xi)^2} \sin^2
\left. \f{(m_3^2 -m_2^2) x}{4 E_{\nu}}
\right] \ .
\label{prob}
\eeq

\noindent (ii) {\it  b = 0}

In this case we obtain with the same parameterization
\beq
c_1 = \frac{\sqrt{\xi}}{\sqrt{1+\xi}}, \; \; \; \; \; \;
s_1 = \frac{1}{\sqrt{1+\xi}}\ ,
\eeq
\beq
m_{eff}^{-1} =
\left (
\begin{array}{ccc}
\frac{1}{m_1} & 0   &    0 \\
0 & 0 & \frac{\sqrt{\xi}}{m_2} \\
0 & \frac{\sqrt{\xi}}{m_2} & \frac{1-\xi}{m_2}
\end{array} \right)
\eeq
and again the expression (\ref{eq:wink}) for $sin^22\theta_1$.
The oscillation probabilities for $\nu_{\mu} \rightarrow \nu_{\tau}$ and
$\nu_{e} \rightarrow \nu_{\tau}$ remain the same.
For the oscillation $\nu_{e} \rightarrow \nu_{\mu}$ we have to
substitute $\xi \rightarrow \frac{1}{\xi}$.

Let us now compare these two possibilities for textures with the
experimental data.
The atmospheric neutrino data implies via (\ref{at1})
that $\xi$ in between
\beq
\xi_1 = 0.23 \; \; {\rm and} \; \; \xi_2 = 4.4\ . \label{eq:grenzen}
\eeq
Since $\xi_1\xi_2 =1$ the value of $\xi$ selected
merely determines which of the
neutrino masses is heavier, as well as the magnitude of
the masses. Indeed,
from  $m_3^2=\frac{\delta m^2}{1-\xi^2}$
and $m_2^2=m_3^2-\delta m^2$, we observe that,
for a value
$\delta m^2 \approx 0.01$ eV$^2$ as implied by the atmospheric
neutrino data only values of $\xi$ very near unity would give
neutrino masses of order $O(1)$ eV. In particular, one may see that
\beq
m_3 \approx m_2 \approx  1 \; {\rm eV}, \; \; \; {\rm for}\ \; \xi = 0.995\ .
\eeq
Here we should note that
this is found by using the results of \cite{fuk} which are quoted
in the introduction and are stricter
than those of \cite{atmo}. In this last reference,
$\delta m^2$ for $\mu-\tau$ oscillations can be as high as
$0.5$ eV$^2$. In this case one finds e.g.
\beq
m_3 = 1.62 \; {\rm eV}, \; \; \; m_2 = 1.45 \; {\rm eV}, \; \; \;
{\rm for}\ \; \xi = 0.90\ .
\eeq

After accommodating the atmospheric neutrino data, we
turn to the discussion of the solar neutrino numbers,
and in this example we interpret them as
$\ne \rightarrow \nm$ oscillations.
{}From (\ref{prob}) we may obtain an effective
$\sin^{2} 2 \theta_{e \mu}$. Depending on the size of $\xi$, the
$\frac{1}{1+\xi}$ or $\frac{\xi}{1+\xi}$ term dominates.
\beq
\xi \ll 1 :\ \  \sin^2 2\theta_{e\mu} \approx \sin^22\theta_{e }
\frac{1}{1+\xi} \approx 1.6 \cdot 10^{-2} \ ,
\eeq
\beq
\xi \gg 1 :\ \  \sin^2 2\theta_{e\mu} \approx \sin^22\theta_{e }
\frac{\xi}{1+\xi} \approx 1.6 \cdot 10^{-2} \ ,
\eeq
when inserting the value of $\theta_e$ from (\ref{eq:zahl}) and $\xi$
from (\ref{eq:grenzen}).
This is just in agreement with the MSW solution (\ref{msw}).
To satisfy the mass constraints,
$m_1$ must be nearly equal to
$m_2$.
For an average mass $m_0 \approx 1$ eV,
$\delta m_{12}^2 \approx 2m_0 \mid m_2-m_1\mid \approx 10^{-5} eV^2$
indicates the need for a very big degeneracy.
Such a high degree of degeneracy is extremely hard to explain from an
underlying
theory without fine tuning, unless the masses are forced to such values
by symmetries. In section 6 we are going to show why this is the case.

Finally we want to discuss neutrinoless double $\beta$-decay and the
COBE data.
For the first one, from (\ref{eq:beta}) and (\ref{eq:mixer}), we obtain
\beq
|<m_{\nu_e}>| = |c_e^2 m_1 + e^{i(\lambda_2 - \lambda_1)}
s_e^2 \left (c_1^2 - \frac{s_1^2}{\xi}e^{i(\lambda_3-\lambda_2)}
\right)m_2 |\ ,
\eeq
where $e^{i(\lambda_2 - \lambda_1)}$
is the relative CP eigenvalue of
$\nu_1$ and $\nu_2$ (the masses here are positive).
Taking $\nu_2$ and $\nu_3$ to have the same CP eigenvalues (as already
discussed in section 2), we obtain
\beq
|<m_{\nu_e}>| = |c_e^2 m_1 + e^{i(\lambda_2 - \lambda_1)}
s_e^2 \left (c_1^2 - \frac{s_1^2}{\xi}\right)m_2 |\ .
\eeq
Now we may again study the texture zeroes.
With (\ref{eq:sin}) we get
\beq
|<m_{\nu_e}>| = c_e^2 m_1 \approx m_1 = O(1) \; {\rm eV}
\eeq
which is consistent with the bound (\ref{eq:beta}).
The above
predictions are consistent with the COBE
data, as well,  since the sum of the masses for the parameter
range we indicate,
can be of order a few eV's, as required.
Therefore we conclude that there is no problem in accommodating
the experimental data for the phenomenological texture
zero solutions. An identical situation occurs
when the large mixing which explains the atmospheric neutrino
deficit is in the $1-2$ entries of the neutrino mass matrices.


\section{Derivation of textures from $U(1)$ symmetries }

In section 3, we already gave the motivation for looking for
texture zeroes, arising due to symmetries
in the underlying string or GUT theory. After obtaining a
classification of three exact and the general phenomenological
texture zero solutions in the appendix resp. section 5, we want
to demonstrate how such patterns come about.
Let us consider the possibility of obtaining
the above textures from additional $U(1)$ symmetries,
following from the work of
 Ibanez and Ross (IR) \cite{IR} as well as \cite{DLLRS,LLR}.
We stress again that such additional $U(1)$ symmetries appear
most naturally in string theories (especially at the ``conformal
point'').
The $U(1)_{FD}$ charges assigned to the matter fields
can be found in IR. They are chosen in such a way as to make the
mass matrices symmetric (resp. hermitian).
Moreover, the lighter generation charges are fixed by the need
to have  anomaly cancellation, which is ensured by taking the $U(1)$
to be traceless. Then one obtains the structure
\begin{eqnarray}
m_u\approx \left(
\begin{array}{ccc}
\epsilon^{\mid -4\alpha _1-2\alpha _2\mid } &
\epsilon^{\mid -3\alpha_1\mid } &
\epsilon^{\mid -  \alpha_2-2\alpha_1\mid }
\\
\epsilon^{\mid -3\alpha_1\mid } &
\epsilon^{\mid 2(\alpha_2-\alpha_1)\mid } &
\epsilon^{\mid \alpha_2-\alpha_1\mid } \\
\epsilon^{\mid -\alpha_2-2\alpha_1\mid } &
\epsilon^{\mid \alpha_2-\alpha_1\mid } & 1
\end{array}
\right)\ ,
\label{eq:mu0}
\end{eqnarray}
which exhibits the relations
\begin{equation}
m^u_{11}\ \simeq \ {{(m^u_{13})^2}\over {m^u_{33}}}\ \ ;\ \
m^u_{22}\ \simeq \ {{(m^u_{23})^2}\over {m^u_{33}}}\ .
\label{eq:gen}
\end{equation}
This structure is consistent with solutions 1, 2 and 4
of the textures shown in table 1. This is because a texture
zero in the (1,3) position is correlated
with a texture zero in the (1,1) position.
In \cite{DLLRS,LLR} a similar analysis had been done
 to derive the Majorana mass matrices from $U(1)$ --
symmetries for the case of the up quark matrix (\ref{eq:mu0}).
In this work, we examined
the simplest case which arises when adding only one new pair of singlet
fields $\Sigma$, $\bar\Sigma$
with zero hypercharge, but charged under the
new $U(1)$ -- symmetry.

Here, we will show how one can derive by symmetries
cases 1 and 3,  which seem to have the optimal structure, especially
for a heavy Majorana mass matrix with many texture zeroes,
as we can see from tables
2 and 3.
In \cite{IR}, the correct $u$-quark mass matrix
is found by making the choice of $\alpha_2/\alpha_1$,
which generates the right order for the non-zero elements
of the solutions 1, 2 or 4. By demanding that
the powers of the
(1,2) and (2,3) matrix elements be in the ratio
3:1 (as needed for solution 2 or 4), $\alpha_2=2\alpha_1$ and
the $u$-quark mass matrix has the form
\begin{eqnarray}
m_u\approx \left(
\begin{array}{ccc}
\epsilon^8 & \epsilon^3 & \epsilon^4 \\
\epsilon^3 & \epsilon^2 & \epsilon \\
\epsilon^4 & \epsilon & 1
\end{array}
\right) \ .
\label{eq:mu}
\end{eqnarray}

Here one also uses the freedom to set $\alpha_1=1$ through a
redefinition of the parameter $\epsilon$ and $\alpha_2$ (i.e.
$\epsilon\rightarrow (\epsilon)^{\alpha_1}, \; \alpha_2
\rightarrow \alpha_2/\alpha_1)$.
Notice that e.g. in solution 2 the (2,2) element is zero,
but it actually does not affect the phenomenology,
if it is up to $\epsilon^2$. Further study reveals that
one may obtain exactly the same hierarchy structure for $m_d$,
where one encounters only a different parameter $\bar\epsilon$
\cite{IR}.
These two matrices closely resemble case 2 in table 1 and are
in agreement with the data.

For possible choices of the lepton masses, we refer the reader
to \cite{IR} and the extension in \cite{DLLRS}. We remark that
$m_u \sim m_{\nu}^D$ is more or less the simplest choice for
the Dirac neutrino masses. But what are the predictions for
the Majorana neutrino mass matrix? The most obvious choice
leads to the same charge pattern as for the $u$-quarks,
with an additional complication coming from the
presence of a singlet field.
As we have already mentioned, right-handed fields
get Majorana masses from a term of the form
$\nu_R\nu_R\Sigma$  where $\Sigma$ is a $SU(3)\otimes
SU(2)\otimes U(1)$ invariant Higgs scalar field with $I_W=0$ and
$\nu_R$ is a right-handed neutrino.
If we assume a $\Sigma$ field with charge $-1$, it will
make the (2,3) entry of the resulting mass matrix
1. Indeed, what we obtain in terms of
$\bar{\epsilon}$ is \cite{DLLRS,LLR}
\begin{eqnarray}
M_{\nu_R}\approx \left(
\begin{array}{ccc}
0 & \bar{\epsilon}^{(-3-1)} & \bar{\epsilon}^{(-4-1)} \\
\bar{\epsilon}^{(-3-1)} & \bar{\epsilon} & 1 \\
\bar{\epsilon}^{(-4-1)} & 1 & \bar{\epsilon}^{(-1)}
\end{array}
\right)\ ,
\label{eq:mumu}
\end{eqnarray}
where we have
set the smaller entry to zero and have not yet taken
the absolute values of the charges in the exponents, since at the next
stage we are going
to introduce a second singlet field,
which alters the structure of the
heavy Majorana mass matrix and
we want to have the effect of the charge of each field manifest.
The matrix in eq.(\ref{eq:mumu}) has
similar structure to the one we derived
in (\ref{eq:heavy}), for the 2-3 sector
(that is the down right $2 \times 2$ submatrix).
To obtain the desired structure for the complete matrix,
we add a second
$\Sigma^\prime$ field which develops a
similar VEV and has a quantum number $+2$ under the $U(1)$ symmetry.
This means that now, in the heavy
Majorana neutrino mass matrix, the dominant element
will be the one with the biggest absolute power in $\bar{\epsilon}$.
I.e., the elements (2,2), (2,3) and (3,3) would still arise mainly
due to the couplings to the
$\Sigma$ field with charge -1, while the (1,2)  and (1,3)
elements arise from the couplings to
to $\Sigma^\prime$. Then the complete matrix is
\begin{eqnarray}
M_{\nu_R}\approx \left(
\begin{array}{ccc}
0 & \bar{\epsilon} & \bar{\epsilon}^{2} \\
\bar{\epsilon} & \bar{\epsilon} & 1 \\
\bar{\epsilon}^{2} & 1 & \bar{\epsilon}
\end{array}
\right)
\label{eq:mumumu}
\end{eqnarray}
and the structure would be that of the example in section 4.
Actually, this is in fact the solution with only
$c=0$ (where the (2,2) element is of order $\bar{\epsilon}$).

Is that generic, in the sense that we may create any mass matrix in that way?
Within the simple procedure of adding only a $U(1)$ symmetry and more
singlet fields, the answer is probably negative.
Nevertheless, going beyond the simple descriptions given above,
while assuming more than one $U(1)$ symmetries the phenomenologically
viable Majorana mass matrices obtained in this work may be derived
naturally.

After examining how the structure of the
heavy Majorana mass matrix may arise, we come back to the
generation of the quark mass matrices for the
prefered cases 1 and 3.
We start with the $u$-quark mass matrix for
the two cases. In case 1 we need the (2,3) element
to be zero. This can be done by assuming that the total charge
of this entry is half--integer and therefore gets banned by a $Z_2$
symmetry. In principle, we could choose a large $U(1)$--charge,
for this entry, which would make it small. However, the
(2,2) entry is the square
of the (2,3) entry as we see from (\ref{eq:gen}).
Therefore the first choice is the correct one.
The (1,2) entry, which is $\epsilon^{|-3\alpha_{1}|}$,
has to be non-zero. Thus $\alpha_1$ is integer,
$\alpha_2$ half-integer.
This implies that not only the (2,3) entry is zero, but also the
(1,3).
The form of the mass matrix is
\begin{eqnarray}
m_u\approx \left(
\begin{array}{ccc}
\epsilon^{\mid -4\alpha _1-2\alpha _2\mid } &
\epsilon^{\mid -3\alpha_1\mid } &
0 \\
\epsilon^{\mid -3\alpha_1\mid } &
\epsilon^{\mid 2(\alpha_2-\alpha_1)\mid } &
0 \\
0 & 0 & 1
\end{array}
\right)\ .
\label{eq:mu1}
\end{eqnarray}
Setting $\alpha_{2} = \frac{w}{2} \alpha_1$,
$w$ = odd integer,
one may choose $w$ so as to have the (1,1) entry in
high power and thus effectively zero, while getting the hierarchical
structure between the two non--zero entries
(1,2) and (2,2).
The relation of the above $\epsilon$ to the one in \cite{IR} and
\cite{DLLRS} should be clear form the context.

Case 3 (the Giudice ansatz) may occur in the following way.
We need the (1,2) entry to be 0, thus it is
$\alpha_1$ which we take to be half-integer.
Taking $\alpha_2$ integer, the (2,3) entry is automatically zero and the mass
matrix is of the form
\begin{eqnarray}
m_u\approx \left(
\begin{array}{ccc}
\epsilon^{\mid -4\alpha _1-2\alpha _2\mid } &
0 &
\epsilon^{\mid -  \alpha_2-2\alpha_1\mid }
\\
0 &
\epsilon^{\mid 2(\alpha_2-\alpha_1)\mid } &
0 \\
\epsilon^{\mid -\alpha_2-2\alpha_1\mid } &
0 & 1
\end{array}
\right)\ .
\label{eq:mu2}
\end{eqnarray}
The entry (1,1) is once more effectively zero,
since it appears at a high power. The (1,3) and (2,2)
entries can be made the same (up to a
coefficient) by setting
\beq
\alpha_2 = 4 \alpha_1\ .
\eeq
After the $u$-quark mass matrices, we have to tackle the
structures for the $d$-quark mass matrices. Here
things are more complicated, but the main idea has already been
given in \cite{7}. One has to use different mixings in the
light Higgses $H_1$, $H_2$. In principle one may create any
structure from a complicated enough mixing. Here we want to
demonstrate that often already simple mixing will do. The general
form is
\begin{equation}
H^{light}_{1,2} = H_{1,2} + \sum_r \left(
 H_{1,2}^r \frac{< \theta >^r}{M_{1,2}^r} +
 H_{1,2}^{-r} \frac{< \bar\theta >^{-r}}{M_{1,2}^{r}} \right)\ ,
\label{eq:higgs}
\end{equation}
where we denote by $H_{1,2}^r$ a Higgs field carrying a $U(1)$ charge
$r$\footnote{Here we assumed for simplicity one pair of fields
$\theta$, $\bar\theta$ with charge $\pm$1. One might also have
several pairs with different charges and couplings.}.
Which elements of a specific mass matrix are actually created
depends entirely on the terms in the sum on the rhs.

In this way, one can reproduce the down quark
mass matrix for Case 1. The (1,3) entry then can be almost
zero, because it can be related to a higher charge, as both
the terms that contain $\alpha_{1,2}$ have the same sign.
However, now, the (2,3) entry is no longer zero.
There seems to be a small problem here. This is
that the (2,2) and (2,3) entries are related, and
while now we get (2,3)$^2 \approx$ (2,2), in the
textures of Ramond-Roberts-Ross
they are of the same order.
However, note that this can be fixed by a choice of coefficients.

Similarly, we can get  Case 3 by a suitable mixing.
The $d$-quark mass matrix here is of the form
\begin{eqnarray}
m_d\approx \left(
\begin{array}{ccc}
0 & \bar{\epsilon}^3 & \bar{\epsilon}^4 \\
\bar{\epsilon}^3 & \bar{\epsilon}^2 & \bar{\epsilon} \\
\bar{\epsilon}^4 & \bar{\epsilon} & 1
\end{array}
\right)\ .
\label{eq:mu4}
\end{eqnarray}
This structure for the down mass matrix is viable.
But the up? We saw before how we can get it in general (\ref{eq:mu2}).
We observe that
the (1,2) and (2,3) entries, which we want to vanish are
odd numbers, while the others that we have to retain are
even. We therefore need a $Z_2$ symmetry to ban the odd
charges in (\ref{eq:mu4}). But this can be done easily, if there are
only fields with even charges in the light Higgs fields (\ref{eq:higgs})
that couples to $m_u$.
Then we get
\begin{eqnarray}
m_u\approx \left(
\begin{array}{ccc}
0 & 0 & \epsilon^4 \\
0 & \epsilon^2 & 0 \\
\epsilon^4 & 0 & 1
\end{array}
\right)
\label{eq:mu5}
\end{eqnarray}
which just gives the (2,2) entry larger than what we would like.
However, the basic structure is the same and coefficients can
make it even better.

\section{Renormalization Effects}

Up to now, in section 4 resp. the appendix
we found all exact three texture zero
solutions at the scale $M_X$, while
in section 5
we discussed phenomenological texture zero solutions at $M_W$ and
found that there is hardly any difficulty in accommodating the
experimental data. From the range of solutions for $\xi$ in
(\ref{eq:grenzen})
it is clear that there
is no problem in reconciling the solutions at both scales.
A natural solution to all the neutrino
puzzles may be obtained when the light effective Majorana
mass scale is of $O(1eV)$. In the context
of a Grand Unified Theory such a small scale is obtained
by the implementation of the {\it see--saw} mechanism
resulting in the effective light (Majorana) mass matrix
$m_{\nu}^{eff}={m_{\nu}^D}^2/M_{\nu_R}$.
As already discussed, in  GUT's the scale $m_{\nu}^D$
is usually fixed by the quark mass matrix
$m_{\nu}^D \sim m_Q$, therefore  the right handed
neutrino scale should be around $M_N\sim 10^{12}-10^{13}GeV$,
i.e. at least three orders smaller
than the gauge unification scale, $M_X\sim 10^{16}GeV$.
Then, the running of the couplings
from the Unification scale, $M_X$,
down to the scale of $M_{\nu_R}$, must include possible
radiative corrections  from  $\nu_R$ neutrinos. After that
scale, $\nu_R$'s decouple from the spectrum, and the effective
{\it see -- saw} mechanism discussed above is operative.
This running - even if of $O(1)$ - will not
be able to spoil the neutrino
hierarchy of the mass matrices. For
the phenomenological zeroes at $M_X$
this is even more obvious.

It has already been observed that the main result of
the presence of $\nu_R$  is a $10\%$ effects in the
$b-\tau$--unification, and that only for small
$\tan\beta$ ($h_t \gg h_b$) \cite{VB}.
It is well known that,  in most of the
unified gauge groups the $b-\tau$ equality is a standard
successful prediction.  Indeed,
after taking into account renormalization group effects
from $M_X$ down to $M_W$, the correct $m_b/m_{\tau}$--ratio
at low energies is obtained naturally, if the Yukawa couplings
$h_b,h_{\tau}$ are equal at the GUT scale.
In the presence of the right--handed neutrino however, the
renormalization group equations (RGE) get modified for
small $\tan\beta$.

Since at this stage we already make a distinction
between small and large $\tan\beta$,
 we should note the following concerning mass matrices:
in the simplest scheme (IR) where one tries to derive
the known fermion masses from $U(1)$ symmetries, the model
is forced to be in the
large $tan\beta$ regime. This is because at the tree level
the $U(1)$ quantum numbers of the light Higgses $H_1,H_2$ allow
them to couple to the third generation and an effective
$SU(2)_l\otimes SU(2)_R$
symmetry of the couplings ensures equal Yukawa couplings
$h_b\approx h_{t}$. Nevertheless the model is easily modified,
if there is an additional heavy state, $H_i,\;\bar
H_i,\; i=1 $ or  2, with the same $U(1)$ quantum number. Then mixing
effects can generate different $h_b$ and $h_{t}$
couplings, allowing for any value of $\tan \beta$.

The RGE for small $\tan\beta$ and for the
third generation Yukawa coupling can be approximated as
\barr
 16\pi^2 \frac{d}{dt} h_t&=
 & \left(
 6 h_t^2  + h_N^2
   - G_U\right)  h_t, \label{eq:rg1}
 \\
 16\pi^2 \frac{d}{dt} h_N&=& \left(
  4h_N^2  + 3 h_t^2
   - G_N \right) h_N,
 \label{eq:rg2}  \\
 16\pi^2 \frac{d}{dt} h_b&=
 & \left(h_t^2 - G_D \right) h_b, \label{eq:rg3}
 \\
 16\pi^2 \frac{d}{dt} h_{\tau}&=&\left( h_N^2
  - G_E \right) h_{\tau},
 \label{eq:rg4}
 \earr
where $h_N$ is the largest Yukawa coupling of the right-handed
neutrinos. The $G_{\alpha}=\sum_{i=1}^3c^i_{\alpha}g_i(t)^2$ are
functions that depend on the gauge couplings and the coefficients
$c_{\alpha}^i$.
Below $M_N$, the right handed neutrino decouples from the
massless spectrum and we are left with the standard spectrum
of the MSSM. Thus for scales $t$ beyond $M_N$ the gauge and Yukawa
 couplings evolve according to the standard renormalization
 group equations. We may see clearly the effect of the
$\nu_R$ threshold on the $b-\tau -$Unification if we
write the relation between the Yukawa couplings at the
$M_{\nu_R}$ scale.
\begin{equation}
 h_{b}(t_N)=\rho
 \xi_t\frac{\gamma_D}{\gamma_E}h_{\tau}(t_N)\ ,  \label{rho}
 \end{equation}
 with $\rho=\frac{h_{b_0}}{h_{\tau_0}\xi_N}$ and
\barr
 \gamma_\alpha(t)&=&  \exp({\frac{1}{16\pi^2}\int_{t_0}^t
  G_\alpha(t) \,dt})\ ,\\
\xi_i&=& \exp({\frac{1}{16\pi^2}\int_{t_0}^t h^2_{i}dt})\ ,
 \earr
where $t_0$ is at the high scale $M_X$.
Here $\xi_i\le 1$.
In the case of $b-\tau$ unification we have
$h_{\tau_0} =h_{b_0}$. Thus in the absence
of the right -- handed neutrino $\xi_N \equiv 1$, which implies
 $\rho =1 $ and the $m_b$ mass has the phenomenologically reasonable
prediction at low energies.
However in the presence of $\nu_R$, if $h_{\tau_0}
=h_{b_0}$ at the GUT scale, the parameter $\rho$
is no longer equal to unity since $\xi_N<1$. In fact the
parameter $\xi_N$ becomes smaller for lower $M_N$ scales.
Therefore in order to restore the correct $m_b/m_{\tau}$ prediction at low
energies we need $\rho =1$, what corresponds to
\begin{equation}
h_{b_0}=h_{\tau_0}\xi_N\ .
\end{equation}

This would seem to alter the relative structure between
the mass matrices, however, there exists a natural
way to retain
the successful $b-\tau$ unification, as it is predicted by
GUT's, with the simultaneous presence of the
desired neutrino mass scale $M_N$ to resolve the neutrino
puzzles.
Such a solution has been proposed in
\cite{LLR} in the context of fermion mass textures predicted
by $U(1)$--symmetries. It was found that it is possible to
retain the $m_b^0=m_{\tau}^0$ GUT prediction of the $(3,3)$ --
elements of the corresponding mass matrices, provided there is
sufficient mixing in the charged lepton mass matrix between
the two heavier generations.
But this mixing is also what one needs in order to
solve the atmospheric neutrino problem.

All this is true for the small $tan\beta$ regime.
In the case of  a large $tan\beta$ a first thing to note is that
there are important corrections to the bottom mass
from one-loop graphs involving SUSY scalar masses and the $\mu$
parameter, which can be of the order of $(30-50)\%$.
Besides this, the effect
of the heavy neutrino scale is much smaller, since now
the bottom Yukawa coupling also runs to a fixed point, therefore
its initial value does not play an important role.
To compare things, we look at the
 maximal possible effect on the
$b-\tau$ unification, which would occur for a scale
$M_N = 10^{12}$ GeV and an upper limit for the running bottom
mass $m_b=4.35$. In this case, for the parameter space where
$h_t = 2.0$ and $h_b= 0.0125$ lead to a factor $\xi_N = 0.86$,
when we set $h_b=2.0$, $\xi_N=0.96$. Moreover, for the same
example, if we allow for a running bottom mass $m_b=4.4$, $\xi_N=0.99$
(remember that the effect of the neutrino is to increase
the bottom/tau mass ratio). For higher heavy neutrino scales,
the relevant effect is even smaller.
However, even for large $\tan\beta$, a strong mixing is
also desired in order to solve the atmospheric neutrino problem.

Finally, an additional effect of renormalization effects,
is that, for large lepton couplings, they
amplify the neutrino mixing angle at the GUT scale, when
going to low energies \cite{tanimoto}.
This is in the correct direction for a solution to the
atmospheric neutrino problem.


\section{ Conclusions}

We have explored the possibility of deriving simple Majorana mass
matrices of right-handed neutrinos, which may explain simultaneously
all the neutrino experimental data (atmospheric neutrino oscillations,
solar neutrino oscillations in the MSW approach, neutrinoless
double $\beta$-decay and the COBE data).
This can be accomplished by assuming the existence
of  a right handed
neutrino Majorana mass matrix $M_{\nu_R}$ with a
scale $(10^{12}-10^{13})$ GeV. The solution of the
atmospheric neutrino puzzle resides in a large mixing stemming from
the neutrino mass matrix. Some type of unification or
partial unification implying $m_{\nu}^D \sim m_u$
was adopted.
This is a common relation in
successful GUTs, since it minimizes the number of arbitrary
parameters and increases the predictability of the theory.

Along these lines, we gave a complete classification of {\it exact}
three zero texture solutions at large scales
$M_X$. These solutions allow just one large
mixing\footnote{We remark that all such solutions are given
in the appendix, even if they do not allow
the required mass degeneracies.}.
On the other hand we studied {\it phenomenological} texture zero solutions
at any scale.
It was found that there is no problem to reconcile both types
of zero solutions with the experimental data\footnote{See also previous
footnote.}.
As we see from table \ref{table:maj2}, in the large $tan\beta$ case,
a natural derivation of the right handed neutrino mass
matrix $M_{\nu_R}$ in terms of the low energy constraints
is obtained for cases 1 and 3.

The inclusion of renormalization
group effects due to the right handed neutrino threshold
does not spoil these observations.
The main effect of including a neutrino running coupling,
is that, retaining the successful $m_b^0=m_{\tau}^0$
prediction at the GUT scale, in the simplest schemes, is
now possible
only in the large $tan\beta$ case. In the small tan$\beta$
scenario, the restoration of $b-\tau$ equality at $M_{GUT}$
requires a large mixing in the charged lepton sector between
the two heavier families, which is sufficient to solve the
atmospheric neutrino puzzle \cite{LLR}.
Interestingly enough, some of the phenomenologically derived
mass textures that are presented,
 can be obtained using additional simple $U(1)$
symmetries along the lines of \cite{IR}, assuming
proper $U(1)$--charges for the standard matter fields and
additional singlets acquiring vacuum expectation values.

\vspace{0.5cm}
\noindent
{\bf  Acknowledgment  }

We would like to thank G. G. Ross for many enlightening
discussions. G. Leontaris and J. Vergados
acknowledge partial support by the CEC project
CHRX-CT93-0132.


\newpage

\begin{center}
{\large {\bf Appendix}}
\end{center}

In this appendix we carry out the systematic study of the three
texture zero solutions that allow one large mixing angle.
In particular, we are interested in three zeroes, since this is in
general the maximally allowed number for three non-zero eigenvalues.
Indeed, although
there exist three cases of matrices with four texture
zeroes and three non-vanishing eigenvalues, applying the
discussion of section 4 to this case leads to an
over-determined set of constraints. Only if there are additional
structures (like the block structure of $m_{\nu}^D$ in case 1 of
table \ref{table:maj}),  solutions exist at all but even in this
case no large mixing may be obtained.

The inverse light neutrino Majorana matrix is
\begin{eqnarray}
m_{eff}^{-1}=\ R \cdot m_{eff}^{-1diag}\cdot R^T\ ,
\label{eq:rot}
\end{eqnarray}
where $R$ are appropriate {\it rotations}.
What type of rotations do we have to study?
Since $m_{\nu}^D \sim m_u$  and
$m_u$ is real and symmetric \footnote{In the framework of
\cite{7}, $m_u$ can
always be chosen to be real symmetric. Possible phases reside
in $m_d$ and are suppressed in table \ref{table:maj}.}, $m_{\nu}^D$ is real
symmetric as well.
$M_{\nu_R}$ and $m_{eff}$ are complex symmetric, and the
complex phases are in general relevant. Trying to absorb them by redefinitions
would let them reappear in $m_{\nu}^D$.
However,  instead of taking the general $R$ required for the diagonalization
of a complex symmetric matrix (Schur rotation), we restrict $R$ to
those diagonalizing a real symmetric matrix {\it and} allow
$m_{eff}^{-1diag}$ to possess negative entries. One may convince
himself that this resumes in general all possible cases. Taking
into account also complex phases will only lead to further constraints
on solutions. Nevertheless we have to stress that only including negative
eigenvalues for $m_{eff}^{-1}$ allows non-trivial solutions.
Thus we only consider
\begin{equation}
R\ =\ R_i\cdot R_j\cdot R_k\ \ , \ \ i,j,k \in 1,2,3,\ not\ equal.
\end{equation}
where $R_i$ denotes a rotation in the $j-k$-plane, with $i\neq j,k$.

We are now looking for all possible three texture
zero solutions that allow at least one large
mixing angle\footnote{Since \begin{equation} m_{eff} = R m_{eff}^{diag}
R^T\ , \end{equation} we have the same $R$ in (\ref{eq:rot}).}.
There are 20 possibilities for three texture zeroes in $M_{\nu_R}$.
Because of the experiment there has to be one large mixing either
in the 1-2 or 2-3 submatrix to explain the atmospheric neutrino data.
We restrict ourselves here to the case of one large mixing and take
the others to be small. (In the actual numerical study nevertheless the
precise formulas have been taken.)
We therefore to solve
\begin{equation}
M_{\nu_R}\ =\ m_{\nu}^{D\dagger}\cdot (R m_{eff}^{-1\ diag} R^T)\cdot
m_{\nu}^D\ ,
\end{equation}
where
\begin{equation}
R = R_1R_2R_3
\label{eq:rotation}
\end{equation}
or permutations with
\begin{equation}
R_1 =
\left (
\begin{array}{ccc}
1 & 0   &    0 \\
0 & c_1 & -s_1 \\
0 & s_1 &  c_1
\end{array} \right)  \ \ ,\ \
R_2 =
\left (
\begin{array}{ccc}
  1 & 0 & -e_2 \\
  0 & 1 &    0 \\
e_2 & 0 &    1
\end{array} \right)  \ \ ,\ \
R_3 =
\left (
\begin{array}{ccc}
  1 & -e_3 & 0 \\
e_3 &    1 & 0 \\
  0 &    0 & 1
\end{array}
\right)
\end{equation}
for the large mixing in the 2-3 submatrix
and its analog for the other case.
The results are given in tables \ref{table:maj4} and \ref{table:maj5}.

Here, one has to note that
the results are dependent
of the order in which we multiply the $R_i$ in (\ref{eq:rotation}). In
general we find that $R=R_1R_2R_3$ exhausts all possibilities and that
any permutation is a subclass. But we will now also see why this
dependence is somehow trivial.
We denote with an asterisk the angles
that are associated
with two degenerate eigenvalues. As explained in \cite{cheng}, one is
able to redefine the physical states in those cases and a mixing
has no physical meaning (as it is e.g. the case for all neutrinos
massless). Solutions with the mixing being undetermined
in this way have been dropped from the tables. The reason for this
is that all solutions with degenerate eigenvalues
only make sense when the texture zeroes are not exact (otherwise
the experimental data cannot be explained), implying by the
smallness of these entries that such an undetermined mixing angle will
give a negligible effect.
\begin{table}
\centering
\begin{tabular}{|cc|}
\hline
\multicolumn{2}{|c|}
{\bf $M_N$ Matrices for Textures of Dirac mass matrix 1}
\\ \hline
$\left( \begin{array}{rrr}
0 & 0 & c \\
0 & d & e \\
c & e & 0
\end{array} \right)$ &
$c_1^2=\frac{1}{2}$ \, \, \,
$e_2 = 0$ \, \, \,
$e_3 = 0$ \, \, \,
$-m_2=m_3$ \\
\hline
\multicolumn{2}{|c|}
{\bf   $M_N$ Matrices for Textures of Dirac mass matrix 2}
\\ \hline
$\left( \begin{array}{rrr}
0 & b & c \\
b & 0 & e \\
c & e & 0
\end{array} \right)$ &
$c_1^2=\frac{1}{2}$ \, \, \,  $e_2 \neq 0$ \, \, \,  $e_3 = 0$ \, \, \,
$\pm m_1<<-m_2=m_3$
\\ \hline
\multicolumn{2}{|c|}
{\bf  $M_N$ Matrices for  Textures of Dirac mass matrix 3}
\\ \hline
$\left( \begin{array}{rrr}
0 & b & 0 \\
b & 0 & e \\
0 & e & f
\end{array} \right)$ &
$c_1^2=\frac{1}{2}$   \, \, \,
$e_2 = 0$ \, \, \,
$e_3 = 0$ \, \, \,
$-m_2=m_3$
\\ \hline
\end{tabular}
\caption{Possible heavy Majorana  textures for large
$\nu_{\mu}-\nu_{\tau}$ mixing.
An asteriskin the table denotes arbitrary
angles that are trivial, since they are associated to a mixing of
degenerate eigenvalues. If there are only two masses given at the
rhs, this implies that the third one is arbitrary. The sign $<>$
means that solutions with $<$ and $>$ are found.}
\label{table:maj4}
\bigskip
\end{table}

\begin{table}
\centering
\begin{tabular}{|cc|}
\hline
\multicolumn{2}{|c|}
{\bf Heavy Majorana  Textures of Dirac mass matrix 1}
\\ \hline
$ \left( \begin{array}{rrr}
0 & b & 0 \\
b & d & 0 \\
0 & 0 & f
\end{array} \right)$ &
$\begin{array}{llll}
c_3^2=\frac{1}{2}
\; \; \; \; \; \; \;
&  \ast    \; \; \; \; \; \; \;
&  e_2 = 0 \; \; \; \; \; \; \;
&  -m_1=m_2=m_3 \\
\; \; \; \; \; \; \;
&  e_1 = 0 \; \; \; \; \; \; \;
&  \ast    \; \; \; \; \; \; \;
& \phantom{-}m_1=m_3=-m_2 \\
\; \; \; \; \; \; \;
&  e_1 = 0 \; \; \; \; \; \; \;
&  e_2 = 0 \; \; \; \; \; \; \;
&  -m_1= m_2 \\
0<c_3^2<1
\; \; \; \; \; \; \;
&  \ast    \; \; \; \; \; \; \;
&  e_2 = 0 \; \; \; \; \; \; \;
&  -m_1<>m_2=m_3 \\
\; \; \; \; \; \; \;
&  e_1 = 0 \; \; \; \; \; \; \;
&  \ast    \; \; \; \; \; \; \;
&  \phantom{-}m_1=m_3<>\pm m_2\\
\; \; \; \; \; \; \;
&  e_1 = 0 \; \; \; \; \; \; \;
&  e_2 = 0 & -m_1<>m_2
\end{array}$
\\ \hline
$\left( \begin{array}{rrr}
a & 0 & 0 \\
0 & d & 0 \\
0 & 0 & f
\end{array} \right)$  &
$\begin{array}{llll}
c_3^2=\frac{1}{2}
\; \; \; \; \; \; \;
&  \ast  \; \; \; \; \; \; \;
& e_2 = 0  \; \; \; \; \; \; \;
& -m_1 \approx m_2=m_3 \\
\; \; \; \; \; \; \;
& e_1 = 0 \; \; \; \; \; \; \;
& \ast  \; \; \; \; \; \; \;
& \ m_1=m_3\approx -m_2 \\
\; \; \; \; \; \; \;
& e_1=0   \; \; \; \; \; \; \;
& e_2=0 \; \; \; \; \; \; \;
& -m_1 \approx m_2\\
0<c_3^2<1
\; \; \; \; \; \; \;
& \ast    \; \; \; \; \; \; \;
& e_2=0 \; \; \; \; \; \; \;
& -m_1<>m_2=m_3 \\
\; \; \; \; \; \; \;
& e_1=0   \; \; \; \; \; \; \;
& \ast  \; \; \; \; \; \; \;
& \ m_1=m_3<>\pm m_2\\
\; \; \; \; \; \; \;
& e_1 =0  \; \; \; \; \; \; \;
& e_2=0 \; \; \; \; \; \; \;
& \pm m_1<>m_2
\end{array}$
\\ \hline
\\ \hline
$\left( \begin{array}{rrr}
a & b & 0 \\
b & 0 & 0 \\
0 & 0 & f
\end{array} \right)$ &
$\begin{array}{llll}
c_3^2=\frac{1}{2}
\; \; \; \; \; \; \;
& \ast  \; \; \; \; \; \; \;
& e_2=0      \; \; \; \; \; \; \;
& -m_1\approx m_2=m_3\\
\; \; \; \; \; \; \;
& e_1=0      \; \; \; \; \; \; \;
& \ast       \; \; \; \; \; \; \;
& m_1=m_3\approx -m_2\\
\; \; \; \; \; \; \;
& e_1=0      \; \; \; \; \; \; \;
& e_2=0      \; \; \; \; \; \; \;
& -m_1\approx m_2\\
0<c_3^2<1
\; \; \; \; \; \; \;
& \ast       \; \; \; \; \; \; \;
& e_2 =    0 \; \; \; \; \; \; \;
& -m_1<>m_2=m_3 \\
\; \; \; \; \; \; \;
& e_1 =    0 \; \; \; \; \; \; \;
& \ast       \; \; \; \; \; \; \;
& m_1=m_3<>-m_2 \\
\; \; \; \; \; \; \;
& e_1 =    0 \; \; \; \; \; \; \;
& e_2 =    0 \; \; \; \; \; \; \;
& -m_1<>m_2
\end{array}$
\\ \hline
\multicolumn{2}{|c|}
{\bf Heavy Majorana Textures of Dirac mass matrix 2}
\\ \hline
$\left( \begin{array}{rrr}
0 & b & 0 \\
b & d & 0 \\
0 & 0 & f
\end{array} \right)$ &
$0<c_3^2<1$ \, \, \,  $e_1 = 0$ \, \, \,
$e_2= 0$ \, \, \,  $(\pm m_1<>\mp m_2)<<m_3$
\\ \hline
\multicolumn{2}{|c|}
{\bf  Heavy Majorana Textures of Dirac mass matrix 3}
\\ \hline
$\left( \begin{array}{rrr}
a & 0 & c \\
0 & 0 & e \\
c & e & 0
\end{array} \right)$ &
$0<c_3^2<1$ \, \, \,  $e_1 = 0$ \, \, \,  $e_2 = 0$ \, \, \,
$\left\{ \begin{array}{r} m_1<<(-m_2<>m_3) \\
                         (m_1<>-m_2)<<m_3 \\
                         (m_1<>m_3)>>-m_2 \end{array} \right.$
\\ \hline
\multicolumn{2}{|c|}
{\bf  Heavy Majorana Textures of Dirac mass matrix 4}
\\ \hline
$\left( \begin{array}{rrr}
0 & b & 0 \\
b & 0 & e \\
0 & e & f
\end{array} \right)$ &
$0<c_3^2<1$ \, \, \,  $e_1 = 0$ \, \, \,  $e_2 = 0$ \, \, \,
$\left\{ \begin{array}{r} (\pm m_1<>\mp m_2)<<m_3 \\
                         (\pm m_1<>m_3)>>\mp m_2 \\
                         \pm m_1<<(\mp m_2<>m_3) \end{array} \right.$
\\ \hline
$\left( \begin{array}{rrr}
0 & b & 0 \\
b & d & 0 \\
0 & 0 & f
\end{array} \right)$ &
$0<c_3^2<1$ \, \, \,  $e_1 = 0$ \, \, \,  $e_2 = 0$ \, \, \,
$\left\{ \begin{array}{r} (\pm m_1<>\mp m_2)<<m_3 \\
                         (\pm m_1<>m_3)>>\mp m_2 \\
                         \pm m_1<<(\mp m_2<>m_3) \end{array} \right.$
\\ \hline
\end{tabular}
\caption{Possible textures for large $\nu_{e}-\nu_{\mu}$ mixing.}
\label{table:maj5}
\end{table}

Let us now summarize and discuss the results of this classification.
As can be seen from the tables there are several solutions for
different Dirac neutrino mass matrices.
It is clear that the three texture zero solutions in the exact form
({\it exact zeroes}) allow {\it only} one large mixing. All the small
mixings are either zero or trivial. Therefore it is necessary that
the mixing for the solar neutrino resides in $V_{\ell}$ in (\ref{eq:mix}).
Nevertheless if the texture zeroes are assumed to be only of
{\it phenomenological} nature, additional small mixings might be created.
Finally, we point out that one may easily rewrite the found
solutions for $M_{\nu_R}$ in tables \ref{table:maj4}, \ref{table:maj5}
in a form analog to the
solutions of \cite{7}. This had been done e.g. for the example in
section 4 in (\ref{eq:pattern}).
\bigskip

\end{document}